# Collective resonances near zero energy induced by a point defect in bilayer graphene

Jhih-Shih You[1], Jian-Ming Tang[2] and Wen-Min Huang[3,*]

[1] *Department of Physics,*

*Harvard University,*

*Cambridge MA 02138, USA.*

[2] *Department of Physics,*

*University of New Hampshire, Durham,*

*New Hampshire 03824-3520, USA.*

[3] *Department of physics,*

*National Chung-Hsing University,*

*Taichung 40227, Taiwan.*

---

[*] Email: wenmin@phys.nchu.edu.tw

## Abstract

Intrinsic defects give rise to scattering processes governing the transport properties of mesoscopic systems. We investigate analytically and numerically the local density of states in Bernal stacking bilayer graphene with a point defect. With Bernal stacking structure, there are two types of lattice sites. One corresponds to connected sites, where carbon atoms from each layer stack on top of each other, and the other corresponds to disconnected sites. From our theoretical study, a picture emerges in which the pronounced zero-energy peak in the local density of states does not attribute to zero-energy impurity states associated to two different types of defects but to a collective phenomenon of the low-energy resonant states induced by the defect. To corroborate this description, we numerically show that at small system size $N$, where $N$ is the number of unit cells, the zero-energy peak near the defect scales as $1/\ln N$ for the quasi-localized zero-energy state and as $1/N$ for the delocalized zero-energy state. As the system size approaches to the thermodynamic limit, the former zero-energy peak becomes a power-law singularity $1/|E|$ in low energies, while the latter is broadened into a Lorentzian shape. A striking point is that both types of zero-energy peaks decay as $1/r^2$ away from the defect, manifesting the quasi-localized character. Based on our results, we propose a general formula for the local density of states in low-energy and in real space. Our study sheds light on this fundamental problem of defects.

# I. INTRODUCTION

Graphene,[1–6] a sheet of carbon atoms, has prominent potential for building high-speed field-effect transistors,[7–9] owing to its high carrier mobilities for both electrons and holes[10,11] and to the strong field effect in the carrier density. However, the lack of an energy gap near the Fermi level limits the switching ratio between the high and low resistances in a monolayer graphene (MLG).[12] The use of a bilayer graphene (BLG) has been proposed, in which an energy gap can be opened using various means.[13–16] Theoretical studies have shown that an energy gap can be introduced by applying an interlayer bias to a BLG.[17–19] A cornerstone to building field-effect transistors in the graphene framework[20–23] is established through experimental demonstrations of gate tunable BLG devices.[24,25]

Since the transport properties are essentially determined by the density of states near the Fermi level, understanding the effect of defects in low energies in BLG[26–42] is crucial for the fundamental studies and technology applications. Recent investigations demonstrated that defects, such as vacancies or adsorption of adatoms atom,[43–48] can induce pronounced peaks in the LDOS at zero energy in MLG[49–52,54,58] and in BLG.[55] The zero-energy peak originating from such defects in MLG was observed by scanning tunneling microscopy.[43,48]

In contrast to MLG, Bernal stacking BLG[56] has two types of lattice sites, denoted as disconnected sites $A_1/B_2$ or connected sites $B_1/A_2$ (see Fig. 1). Two zero-energy impurity states, associated with the two different positions of vacancies, have been solved analytically[55]: For a vacancy located at a $B_1$ or $A_2$ site, a quasi-localized mode is living in the same layer and exhibiting $1/r$ decay away from the vacancy. For a vacancy located at a $A_1$ or $B_2$ site the zero-energy state behaves quasi-localized in one of the layers where the defect resides and delocalized in the other.[55]

It might be nature to attribute the sharp zero-energy peak in the LDOS near a defect site to a single impurity state at zero energy. However, previous studies in MLG[54,58] have shown that the contribution to the LDOS from the single impurity state at zero energy vanishes in the thermodynamic limit. In particular, the LDOS has a power-law singularity $1/|E|$, which comes from a collective phenomenon of the low-energy resonant states induced by a point defect[54,58]. In this respect, this paper serves to understand the cause for the zero-energy peak in the LDOS in Bernal stacking BLG and to derive analytical expressions for the LDOS in low energies.

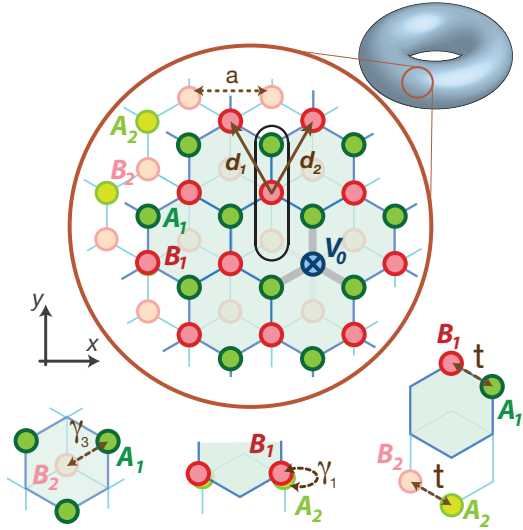

FIG. 1: A nanotorus of BLG with a point defect on the upper layer. The top layer is sketched in a blue shadow, as compared to the bottom one. The black line represents an unit cell, which encloses four sites in sublattices $A, B$ and layers $1, 2$. The intra-layer hopping is denoted as $t$, and two different inter-layer hoppings are respectively represented as $\gamma_1$ between $B_1$ and $A_2$ (connected sites) and $\gamma_3$ between $A_1$ and $B_2$ (disconnected sites), in the unit of $t$. The defect potential is denoted as $V_0$.

We start with a BLG nanotorus in the presence of a point defect, as shown as Fig. 1. At finite system size $N$, the number of unit cells, we numerically compute the LDOS at the nearest-neighbor site of the defect. For a point defect at a connected site $B_1/A_2$, we find that the spectral weight of the zero-energy peak in the LDOS scales as $1/\ln N$. The $1/\ln N$ behavior is attributed to the zero-energy state which is quasi-localized. In additional to the zero-energy state, however, we also find enormous induced resonant states with large spectral weights near zero energy. When the size $N$ approaches infinite in the thermodynamic limit, these resonant peaks crowd to zero energy and the zero-energy peak saturates at a finite value. For a point defect placed at a disconnected site $A_1/B_2$, our numerical results show that the spectral weight of the zero-energy peak scales as $1/N$, which signals the delocalized character of an zero-energy state. When $N$ increases to infinite, the zero-energy peak also saturates at a finite value due to the collective phenomenon of the induced resonant states.

To confirm our finite-size calculation, we use the Green's function techniques to derive analytical expressions for the LDOS in low energies and in the thermodynamic limit. Before and after placing a point defect at a connected site $B_1/A_2$, we find that the change of the LDOS at an adjacent site exhibits $1/|E|$ power-law singularity, similar to our previous finding in MLG[58]. On the other hand, for a defect at a disconnected site $A_1/B_2$, the change of the LDOS behaves as a Lorentzian function. The half-width of the Lorentzian is proportional to the inter-layer hopping amplitude on connected sites, $\gamma_1$, in Fig. 1.

To have a complete understanding of the LDOS, we further study the spatial profile of the zero-energy peak around a point defect. Our numerical calculation shows that the spectral weight of the zero-energy peak decays as $1/r^2$ away from both the point defects located at a $A_1/B_2$ and a $B_1/A_2$ site. The $1/r^2$ dependence implies a quasi-localized character of the collective resonant states. Based on our numerical support, we propose a general formula for the LDOS in low energies and in real space.

## II. RESULTS

### A. Tight-binding Hamiltonian of a BLG

The electronic structure of BLG nanotorus can be captured within a tight-binding approach with periodic boundary conditions. As illustrated in Fig. 1 the tight-binding Hamiltonian retains the hopping terms: $t$ is the intra-layer hopping amplitude between nearest-neighbor sites, and $\gamma_1$ and $\gamma_3$ are the inter-layer hopping amplitudes (in the unit of $t$) between $B_1$ and $A_2$ (connected sites) and $A_1$ and $B_2$ (disconnected sites), respectively. We take units in $\hbar = 1$, $t = 1$ and lattice constant $a = 1$. According to previous studies,[16,55,57] the magnitudes of $\gamma_1$ and $\gamma_3$ are about 0.1. With the unit cell shown in Fig. 1, the tight-binding Hamiltonian for BLG is represented as

$$
\begin{aligned}
H_0 = \sum_{\boldsymbol{r}} \Bigg\{ & \sum_{i=1}^{2} \left[ c_{A_i}^\dagger(\boldsymbol{r}) c_{B_i}(\boldsymbol{r}) + \sum_{j=1}^{2} c_{A_i}^\dagger(\boldsymbol{r}) c_{B_i}(\boldsymbol{r}+\boldsymbol{d}_j) \right] \\
& + \gamma_1 \left[ c_{A_2}^\dagger(\boldsymbol{r}) c_{B_1}(\boldsymbol{r}) \right] + \gamma_3 \Big[ c_{A_1}^\dagger(\boldsymbol{r}) c_{B_2}(\boldsymbol{r}+\boldsymbol{d}_1+\boldsymbol{d}_2) \\
& + c_{A_1}^\dagger(\boldsymbol{r}) c_{B_2}(\boldsymbol{r}+\boldsymbol{d}_1) + c_{A_1}^\dagger(\boldsymbol{r}) c_{B_2}(\boldsymbol{r}+\boldsymbol{d}_2) \Big] \Bigg\} + \text{h.c.}, \quad (1)
\end{aligned}
$$

where $c_s(\boldsymbol{r})$ is a fermion annihilation operator at site $s = (A_1, A_2, B_1, B_2)$ of the unit cell at $\boldsymbol{r}$, and $\boldsymbol{d}_{1,2} = (\mp a/2, \sqrt{3}a/2)$ are the lattice vectors. We place a point defect at site $s$ of the unit cell at the origin $\boldsymbol{r} = \boldsymbol{0}$. The defect is described by the Hamiltonian, $H_I = V_0 c_s^\dagger(\boldsymbol{0})c_s(\boldsymbol{0})$, with an impurity potential $V_0$. We note that a vacancy, which corresponds to the elimination of lattice sites without lattice relaxation, is equivalent to the unitary limit of impurity potential, $V_0/t \to \infty$.

Before we consider the model with a point defect, it is instructive to understand the electronic structure of a pristine BLG in low energy. In the basis of $\Psi^\dagger(\boldsymbol{k}) = [c_{A_1}^\dagger(\boldsymbol{k}), c_{A_2}^\dagger(\boldsymbol{k}), c_{B_1}^\dagger(\boldsymbol{k}), c_{B_2}^\dagger(\boldsymbol{k})]$, the Hamiltonian in the momentum space is represented as $H_0 = \sum_{\boldsymbol{k}} \Psi^\dagger(\boldsymbol{k})\mathcal{H}_0(\boldsymbol{k})\Psi(\boldsymbol{k})$, where the $4 \times 4$ matrix $\mathcal{H}_0$ has the following form,

$$\mathcal{H}_0(\boldsymbol{k}) = \begin{bmatrix} 0 & 0 & h(\boldsymbol{k}) & \gamma_3 \tilde{h}(\boldsymbol{k}) \\ 0 & 0 & \gamma_1 & h(\boldsymbol{k}) \\ h^*(\boldsymbol{k}) & \gamma_1 & 0 & 0 \\ \gamma_3 \tilde{h}^*(\boldsymbol{k}) & h^*(\boldsymbol{k}) & 0 & 0 \end{bmatrix}, \tag{2}$$

with $h(\boldsymbol{k}) = 1 + 2\cos\left(\frac{k_x}{2}\right)e^{i\frac{\sqrt{3}}{2}k_y}$, $\tilde{h}(\boldsymbol{k}) = e^{i\sqrt{3}k_y} + 2\cos\left(\frac{k_x}{2}\right)e^{i\frac{\sqrt{3}}{2}k_y}$. Because of $h(\boldsymbol{k}_D) = \tilde{h}(\boldsymbol{k}_D) = 0$ at Dirac points $\boldsymbol{k}_D = (k_x, k_y) = (\pm 4\pi/3, 0)$, the eigenstates at zero energy are

$$\Psi_{q\pm}(\boldsymbol{k}_D) = \frac{1}{\sqrt{2}}\begin{bmatrix} 1 \\ 0 \\ 0 \\ \pm 1 \end{bmatrix}, \quad \Psi_{g\pm}(\boldsymbol{k}_D) = \frac{1}{\sqrt{2}}\begin{bmatrix} 0 \\ 1 \\ \pm 1 \\ 0 \end{bmatrix}. \tag{3}$$

Here $q\pm$ states correspond to the gapless continuum with quadratic dispersion $E(k) = k^2/(2m^*)$ and $m^* = \gamma_1/2$, and $g\pm$ states correspond to the bands with finite gap $\pm\gamma_1$. It is evident that the $q\pm$ states have large amplitudes at the disconnected sites $A_1/B_2$ and the $g\pm$ states have large amplitudes at the connected sites $B_1/A_2$. After introducing a point defect to a BLG, we can distinguish two different types of LDOS, due to contribution from the gapless continuum or the gapped one, associated with the position of the defect. These will then be investigated analytically and numerically in the following sections.

## B. Finite-size Calculation

For a defect placed at a connected site $B_1$, we study the LDOS at the first-nearest-neighbor $A_1$ site in a reference frame centered at the defect position. The $N$ dependent spectral weight of the LDOS represents the intriguing localization character of defect-induced states. First let us consider $\gamma_3 = 0$. As illustrated in Fig. 2 our numerical results reveal that the height of the peak at zero energy scales as $1/\ln N$ when $N$ is smaller than $10^6$. The $1/\ln N$ behavior is a consequence of the quasilocalized zero-energy state existing around the point defect, which was discussed in the previous study[55]. However, as $N$ increases to $N \sim 10^7$, the spectral weight would eventually saturate at a finite value. For $\gamma_1 = \gamma_3 = 0.1$, as shown in Fig. 2, the weight of the zero-energy peak decreases with strong oscillations. Nevertheless, the spectral weight still saturates at a finite value near $N \sim 10^7$.

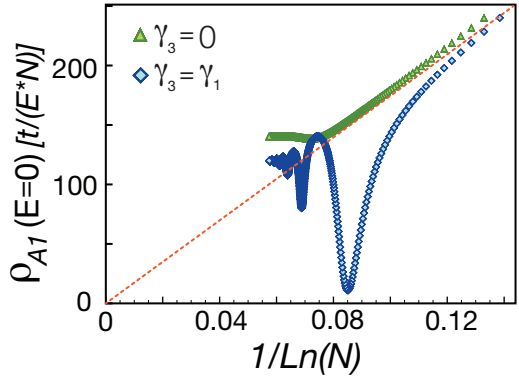

FIG. 2: For a point defect at a $B_1$ site, we illustrate the spectral weight of the zero-energy peak in the LDOS of the nearest-neighbor $A_1$ site versus the size $N$. Blue diamonds and green triangles are represented the data for $\gamma_1 = 0.1, \gamma_3 = 0$ and for $\gamma_1 = \gamma_3 = 0.1$, respectively. The dashed line is a guide to the eye.

To understand the saturation, one might observe the low-energy behavior of the LDOS with respect to different system sizes. Similar to previous discovery in a MLG[58], we find that a point defect also generates a lots of resonant peaks with large spectral weights near zero energy. When the system size approaches to infinity, the defect-induced resonant peaks crowd to zero energy, and eventually the spectral weight at zero energy saturates. The collection of these resonant states constitutes the zero-bias anomaly in the LDOS. In Sec. II C, we will analytically compute the LDOS in low-energy and in the thermodynamic limit and show that the peak is a power-law singularity.

For a defect placed at a connected site $A_1$, Fig. 2 shows the $N$ dependent zero-energy peak of LDOS at the $B_1$ site nearest to the defect. For $\gamma_3 = 0$ and $N < 10^6$, the height of zero-energy peak scales as $1/N$, which is attributed to a delocalized zero-energy state induced by a defect at a $A_1/B_2$ site.[55] When $N$ is larger than $10^6$, the spectral weight at zero energy saturates at a finite value. By investigating the low-energy behavior of the LDOS with respect to different system sizes, we find that the defect-induced resonant states crowd into the zero-energy regime when $N$ increases to infinity. When $\gamma_1 = \gamma_3 = 0.1$, the height of the zero-energy peak decreases with oscillations, as shown in Fig. 2. Nevertheless, the spectral weight still saturates at a finite value.

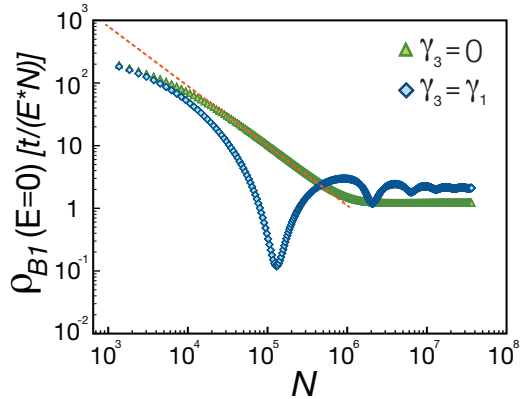

FIG. 3: For a point defect at a $A_1$ site, the spectral weight of the zero-energy peak in the LDOS at the nearest-neighbor $B_1$ site are presented versus the size $N$. Blue diamonds and green triangles are denoted the data for $\gamma_1 = 0.1, \gamma_3 = 0$ and for $\gamma_1 = \gamma_3 = 0.1$, respectively. The dashed line with a slope $-1$ is a guide to the eye.

We emphasize that the spectral weights of the induced resonant states are much smaller than those around a defect at $B_1/A_2$ site. Therefore the height of the zero-energy peak is saturated at a relatively small value, as shown in Fig. 3. In the following, we will analytically compute the LDOS in the low-energy and the thermodynamic limit.

## C. Analytical Computation in the Thermodynamic Limit

In previous section, we focused on an exact numerical evaluation of the LDOS in finite-size systems. For small $N$ the finite-size-scaling of the zero-energy peak follows the localization character of zero-energy states induced by a point defect. When $N$ becomes large, collective phenomena from the induced resonant states near zero energy are expected to become of particular relevance. Eventually the zero-energy peak does not vanish but saturates at a finite value. Now we will analytically derive the change of the LDOS in the low-energy and thermodynamic limit.

The Green's function techniques allow us to obtain the change of the LDOS

$$\Delta\rho_i = -\frac{1}{\pi}\text{Im}\left[\frac{V_0 G_{ij}G_{ji}}{1 - V_0 G_{jj}}\right] \simeq \frac{1}{\pi}\text{Im}\left[\frac{G_{ij}^2}{G_{jj}}\right], \tag{4}$$

at $i$ site nearby the defect at $j$ site, where $i, j = A_1, B_1, A_2, B_2$ are within the same unit cell. In the last step, we used time-reversal symmetry and took the unitary limit $V_0/t \to \infty$, which makes $\Delta\rho_i$ independent on the strength of the impurity potential.

We are interested in the low-energy regime near the Dirac points $\boldsymbol{k}_D = (\pm 4\pi/3, 0)$, where the Hamiltonian, Eq. (2), is written as

$$\mathcal{H}_0 \simeq \begin{bmatrix} 0 & 0 & h(\boldsymbol{q}) & 0 \\ 0 & 0 & \gamma_1 & h(\boldsymbol{q}) \\ h^*(\boldsymbol{q}) & \gamma_1 & 0 & 0 \\ 0 & h^*(\boldsymbol{q}) & 0 & 0 \end{bmatrix}, \tag{5}$$

with $\boldsymbol{q} = \boldsymbol{k} - \boldsymbol{k}_D$. Here $\gamma_3\tilde{h}(\boldsymbol{q})$ is neglected because in the low-energy regime $|\gamma_3\tilde{h}(\boldsymbol{q})| \ll \gamma_1, |\tilde{h}(\boldsymbol{q})|$. The Hamiltonian, Eq. (5), is diagonalized in the eigenbasis, $\Phi^\dagger(\boldsymbol{q}) = \left(\phi_{g-}^\dagger(\boldsymbol{q}), \phi_{q-}^\dagger(\boldsymbol{q}), \phi_{q+}^\dagger(\boldsymbol{q}), \phi_{g+}^\dagger(\boldsymbol{q})\right)$, where $q\pm$ correspond to the quadratic bands $\epsilon_{q\pm} = \pm q^2/\gamma_1$, and $g\pm$ correspond to the gapped bands $\epsilon_{g\pm} = \pm\gamma_1$. Evaluating the Green's functions analytically (for details see **Method**), we obtain the energy dependance of LDOS for two different types of defects.

### 1. A Point Defect at a Connected Site $B_1/A_2$

To the leading order, the retarded Green's functions are approximated as $G_{B_1 B_1} \sim E \ln|E| - i|E|$ and $G_{A_1 B_1} \sim -\Lambda^2 - iE$, where $\Lambda$ is a high-momentum cut-off. If a point defect is placed at a connected $B_1$ site, the change of the LDOS at the nearest-neighbor $A_1$ site is approximated as

$$\Delta\rho_{A_1} \simeq \frac{1}{\pi}\text{Im}\left[\frac{G_{A_1 B_1}^2}{G_{B_1 B_1}}\right] \sim \frac{\Lambda^2}{|E|(\ln|E|)^2}. \tag{6}$$

To confirm our analytical result, we compute the LDOS numerically in the thermodynamic limit. In the left panel of Fig. 4, we show the LDOS of a $A_1$ site before and after placing a point defect at the nearest $B_1$ site. As one can discern from Fig. 4, there exists large spectral weight transferred from the high-energy regime to the low-energy one in the presence of a point defect. Because the logarithmic correction is very weak, our numerical results, shown in the right panel of Fig. 4, exhibit a $1/E$ power-law singularity for the change of the LDOS. We find that the power-law singularity are robust for both $\gamma_1 = \gamma_3 = 0.1$ and $\gamma_1 = 0.1, \gamma_3 = 0$. These results on the power-law singularity allow us to draw connections to the previous study in MLG[58]. We will develop a simple interpretation in terms of Harper equations in the subsection II C 3.

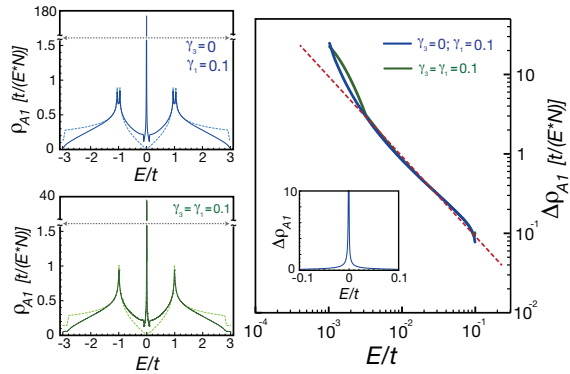

FIG. 4: In the left panels, the solid lines are the LDOS an the $A_1$ site when a point defect is placed at the nearest-neighbor $B_1$ site. In contrast, the dashed lines are denoted as the LDOS for BLG without point defect. In the right panel, we plot the change of the LDOS in the log-log plot. The red dashed line with slope $-1$ is a guide to the eyes.

### 2. A Point Defect at a Disconnected Site $A_1/B_2$

When a point defect is located at a $A_1$ site, we approximate the retarded Green's function to the leading order, $G_{A_1 A_1}(E) \sim E - i\gamma_1$, as shown in **Method**. Thus, the change of the LDOS at the nearest-neighbor $B_1$ site is expressed as

$$\Delta \rho_{B_1} \simeq \frac{1}{\pi} \text{Im} \left[ \frac{G_{A_1 B_1}^2}{G_{A_1 A_1}} \right] \sim \left( \frac{\gamma_1}{E^2 + \gamma_1^2} \right) \Lambda^4. \tag{7}$$

It is remarkable that the change of the LDOS takes the form of Lorentzian function with the broadening factor $\gamma_1$, the inter-layer hopping. To confirm the analytical result, we show the numerical study of the LDOS in the thermodynamic limit in Fig. 5. The changes of the LDOS, shown in the right panel of Fig. 5, are consistent with Eq. (7). A simple interpretation for the Lorentzian broadening will be given in terms of Harper equations in the subsection II C 3.

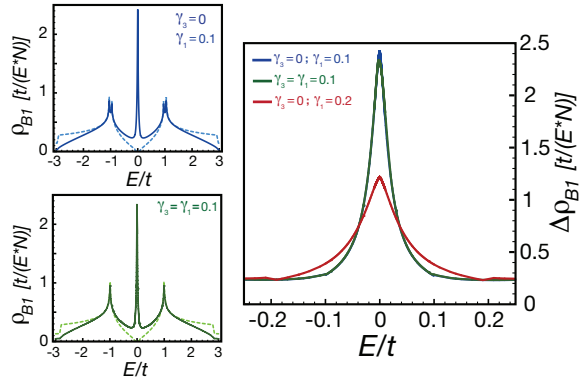

FIG. 5: In the left panels, the solid lines are the LDOS at the $B_1$ site when a point defect is placed at the nearest-neighbor $A_1$ site. In contrast, the dashed lines are denoted as the LDOS for BLG without point defect. In the right panel, we illustrate the change of the LDOS with different parameters in the vicinity of zero energy.

### 3. Analysis of Harper equations

The two different types of the LDOS can be understood from an analysis of Harper equations (See the Harper equations in **Method**). Because $V_0/t \gg 1 \gg \gamma_1, \gamma_3$, it is reasonable to ignore the inter-layer hopping $\gamma_1, \gamma_3$ momentarily. Therefore, the problem is reduced to a point defect problem in MLG. As the system size grows to infinity, the zero-energy singularity[54,58] induced by a point defect at $A_1$ or at $B_1$ in MLG suggests the $E = 0$ state being $\Psi_0(\boldsymbol{r}) = [0, 0, \varphi_{B_1}(\boldsymbol{r}), 0]^T$ or $\Psi_0(\boldsymbol{r}) = [\varphi_{A_1}(\boldsymbol{r}), 0, 0, 0]^T$, respectively.

For a point defect at a $B_1$ site, the Harper equations involving $\varphi_{A_1}$ do not change at all if we turn on the inter-layer hopping $\gamma_1$ but keep $\gamma_3 = 0$. Instead, turning on $\gamma_3$ will cause the wave function $\varphi_{A_1}$ spreading to $A_2$ sites. The effect is described by the Harper equation

at $B_2$ sites:

$$\sum_{\boldsymbol{\delta}_i'} \varphi_{A_2}(\boldsymbol{r} + \boldsymbol{\delta}_i') + \gamma_3 \sum_{\boldsymbol{\delta}_i} \varphi_{A_1}(\boldsymbol{r} + \boldsymbol{\delta}_i) = 0, \tag{8}$$

where $\boldsymbol{\delta}_i$ and $\boldsymbol{\delta}_i'$ are the three displacement vectors pointed from a $B_2$ site to the nearest $A_1$ sites on the top layer and to the nearest $A_2$ sites on the bottom layer, respectively. Because of $\gamma_3 \ll 1$, the spatial wave function $\varphi_{A_1}(\boldsymbol{r})$ remains more or less robust, and $\varphi_{A_2}(\boldsymbol{r})$ is of the order of $\gamma_3$. Thus, $\varphi_{A_2}(\boldsymbol{r})$ accounts for a rather small spread of the wave function from $A_1$ sites to $A_2$ sites. Since in the low-energy limit excitations living on $A_2$ sites are gapped, there is no significant effect for the zero-energy impurity states being coupled to a gapped continuum. Therefore, the LDOS in low energies can be understood in a simple monolayer picture, where the gapless continuum living on $A_1$ sites is disturbed by a point defect. Based on the previous study of MLG[58], significant defect-induced resonant states from the continuum lead to a power-law singularity in the LDOS. Our numerical and analytical results indeed confirm our argument in the thermodynamic limit for BLG.

Now we consider the case of a point defect at a $A_1$ site. When we gradually turn on the inter-layer hopping $\gamma_1$, this gives rise to the Harper equation of $A_2$ site as

$$\sum_{\boldsymbol{\delta}_i} \varphi_{B_2}(\boldsymbol{r} + \boldsymbol{\delta}_i) + \gamma_1 \varphi_{B_1}(\boldsymbol{r}) = 0. \tag{9}$$

where $\boldsymbol{\delta}_i$ represents the set of the three displacement vectors pointed from a $A_2$ site to the nearest $B_2$ sites on the top layer. Following the same argument, $\varphi_{B_2}(\boldsymbol{r})$ is of the order of $\gamma_1$. The hopping amplitude $\gamma_1$ can be viewed as a coupling between the zero-energy state living on $B_1$ sites and the gapless continuum on $B_2$ sites of the bottom layer. It explains that a sharp delta-function peak from a single state in the spectral function is broadened into a Lorentzian shape when coupling to a continuum. Meanwhile, the half-width of the Lorentzian is proportional to the coupling, the interlayer hopping amplitude $\gamma_1$. This is indeed what happens in the our numerical and analytical results.

### D. The spatial profiles of the zero-energy peak.

While the numerical and analytic results in previous sections focused on LDOS in the energy domain, now we study the zero-energy peak in the spatial domain. In Fig. 6, we identify that the spatial dependance of the spectral weight at zero energy is proportional to $1/r^2$ for two different types of point defects, located at $A_1$ or $B_1$ site. This $1/r^2$ behavior manifests the quasi-localized character of the zero-energy peak. In previous sections we have excluded a single impurity state, either quasi-localized or extended, at zero energy as the cause for those peak in the thermodynamic limit. Figure 6, however, suggests that the induced resonant states which crowd to zero energy share the same spatial profile with the quasi-localized state.

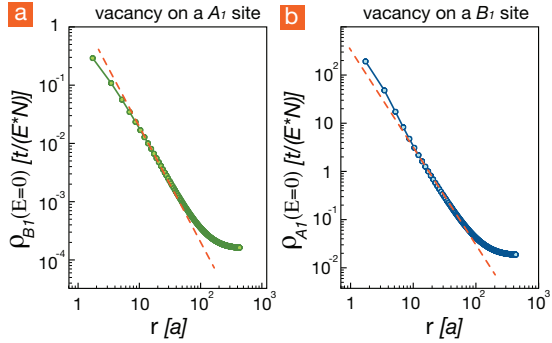

FIG. 6: The spectral weight of the zero-energy peak at (a) $B_1$ sites and (b) $A_1$ sites as a function of the distance $r$ in a reference frame centered at the point defect placed at $A_1$ and $B_1$, respectively. Here we consider the BLG in the thermodynamic limit. The red dashed lines with slope $-2$ are guides to the eyes.

Although the spatial dependence of the resonant states is inaccessible by analytic approach, our numerical support allows us to propose an asymptotic formula for the LDOS in low energies

$$\rho_i(\boldsymbol{r}, E) \simeq F(\boldsymbol{r}) \left[ \frac{1}{C_i} \delta(E) + \Delta\rho_i(E) \right], \tag{10}$$

where $i = A_1, B_1$. The factor $F(\boldsymbol{r}) = 1/r^2$ shows the spatial profile with the quasi-localized character. The first term represents the quasi-localized zero-energy state with the normalization $C_{A_1} = \ln N$ for a defect at a connected site $B_1$, or the delocalized one with the normalization $C_{B_1} = N$ for a defect at a disconnected site $A_1$.[55] The second term $\Delta\rho_i(E)$, defined by Eq. (6) or (7), is the contribution from the resonant states in the low-energy regime induced by the two different types of defects. We note that here the LDOS from a pristine BLG is neglected, because it is much smaller than above two terms. The LDOS in Eq. (10) shows that as the system size approaches to the thermodynamic limit, the contri-bution from zero modes fades away and the LDOS is dominated by infinite resonant peaks. We emphasize that the formula, Eq. (10), is proposed based on our numerical observation. To fully understand the LDOS in low energies, analytic solutions for the resonant states are still necessary.

## III. DISCUSSIONS

We have studied the zero-energy peak induced by a point defect in a Bernal stacking BLG. We numerically computed the LDOS at the first-nearest-neighbor of a point defect and investigated the system-size $N$ dependence of the LDOS. For small size, the zero-energy peak of the LDOS scales as $1/\ln N$ for the induced quasi-localized state or as $1/N$ for the delocalized state at zero energy. When $N$ approaches infinity, the defect-induced resonant states crowd into zero energy and lead to non-vanishing zero-energy peaks for both cases. To further support our numerical findings, we analytically evaluated the change of the LDOS in the thermodynamic limit. We found that the zero-energy peak for a defect at a $B_1/A_2$ site becomes a power-law singularity, while the peak for a defect at a $A_1/B_2$ site is broadened into a Lorentzian shape. By studying the spatial dependence of the zero-energy peak, we showed that the zero-energy peaks for both cases decay as $1/r^2$ away from the point defect. Combing all above discoveries, we proposed a formula for the LDOS in low energies and real space.

The previous theoretical study in monolayer graphene shows that a finite density of vacancies leads to a sharp peak of LDOS exactly at the Fermi level, superimposed upon the flat portion of the DOS[49]. This impurity band can be observed from the numerical calculations. The quasi-localization nature of defect states enables an impurity band form near zero energy, and it is shown that defects induce an impurity band with density of state characterized by the Wigner semi-circle law[58]. The band width of the impurity band is proportional to the square root of the density of defects. Since the impurity band is a flat-band, which supports ferromagnetism when electrons correlation effect is included[59]. In BLG, we expect that an impurity band will appear with finite defect concentration, and the peak of the impurity band will still pin at the Fermi level. The quasi-localization nature of LDOS in BLG discussed in our study may, although further study is needed, exhibit similar physics which a monolayer graphene has.

In the present work we do not investigate correlation effects in the problem of BLG with a monovacancy. Here we would like to discuss the possibility of defect-induced magnetism due to the Coulomb interaction and the large enhancement of the local density of states near the vacancy. By including inter-electronic interactions on a pure $\pi$-band model of BLG, the enhanced LDOS around impurity sites implies that defects may lead to the formation of local magnetic moments[60]. Eventually the spin-split DOS could be observed and the peak of the enhance LDOS will not exactly locate at the Fermi level. However, in real graphene we should consider the $sp^2$ $\sigma$-orbital electrons and the lattice reconstruction near vacancies. To understand a reconstructed single-atom vacancy in graphene, we shall rely on the results from first-principle calculations[61,62].

In principle, near a single atom vacancy there are three unsatisfied $sp^2$ $\sigma$-orbital electrons and one $\pi$-orbital electron. To maximize its spin configurations, a net magnetic moment is $4 \mu B$. If the dangling $\sigma$ orbitals are not passivated, the vacancy reconstructs. If we consider a planar structure, $\pi$ and $\sigma$ bonds do not mix. The *ab initio* calculations have showed that these three impurity levels from the three dangling $\sigma$ bonds split due to the crystal field and a Jahn-Teller distortion[61,62]. The $\pi$ bond state, however, remains at the Fermi energy, being introduced in the midgap of the $\pi$ bands, which refers to a zero-energy impurity state[62]. If the relaxed structure for the vacancy allows non-coplanar structures with out-of-plane displacements, the $\sigma$ orbitals near the Fermi energy will be able to hybridize with the $\pi$ band. Since the zero-energy state decays as $1/r$, its overlap with $\sigma$ orbital states is large. This produces the dominant exchange interaction from Coulomb interaction between local zero-energy state from the $\pi$ orbital electron and the $\sigma$ orbital electron near the Fermi energy. This exchange, due to overlap between the $\pi$ impurity state with the $\sigma$ orbital state, is responsible to the spin-splitting of the vacancy-induced zero-energy $\pi$ states. However, in the literature, the predicted magnetic moment varies widely for the defect[61–68]. The formation of the enhanced LDOS at Fermi level is crucial for determining the hybridization function between the impurity and itinerate states in the picture of the Anderson-Kondo model[69–71]. For the bilayer graphene with Bernal stacking, the situation is more involved since there are two types of vacancies with different LDOS.

The hybridization between the local magnetic moment and the $\pi$ conduction band is proportional to the LDOS. In MLG, the LDOS, $\frac{1}{|E|(\ln |E|)^2}$, leads to the large enhancement of the Kondo temperature and several new types of impurity phases[70]. For BLG, further investigations will be on exploring how the two types of zero-energy peaks evolve when electronic correlations being considered. Since the correlation effects may be significant near defects, our study on LDOS here becomes important on the idea of fabricating spin qubits by defects[72]. The inclusion of realistic interactions to defect-induced resonant states is expected to become of particular relevance for maintaining quantum coherence for a qubit. Moreover, recent experiments [74] showed that in graphene localized states induced by disorders can significantly enhance electron-phonon coupling and become a local drain of hot carriers, which is crucial for electronic transport at low temperatures[73]. For BLG, two types of defects could play different roles on dissipation. The thermal imaging of dissipation shall be revealed by a superconducting quantum interference nano-thermometer[74].

## IV. METHOD

### A. Green's function

To calculate the LDOS, we solve the Dyson equation for BLG in the presence of a point defect with a potential $V_0$. The LDOS can be expressed in terms of the non-interacting retarded Green's function. Assuming the defect placed at $j$ site of the unit cell at the origin $\boldsymbol{r} = \boldsymbol{0}$, we use standard Green's function techniques to formulate the LDOS at $i$ site of the unit cell at $\boldsymbol{r}$ as

$$\rho_i(E, \boldsymbol{r}) = \rho_i^0(E, \boldsymbol{r}) + \Delta\rho_i(E, \boldsymbol{r}). \tag{11}$$

Here the LDOS and the change of the LDOS are represented respectively as

$$\rho_i^0(E, \boldsymbol{r}) = -\frac{1}{\pi} \text{Im} \left[ G_{ii}(E, \boldsymbol{r}) \right], \tag{12}$$

$$\Delta\rho_i(E, \boldsymbol{r}) = -\frac{1}{\pi} \text{Im} \left[ \frac{V_0 G_{ij}(E, \boldsymbol{r}) G_{ji}(E, -\boldsymbol{r})}{1 - V_0 G_{jj}(E, \boldsymbol{0})} \right], \tag{13}$$

with $i, j = A_1, A_2, B_1, B_2$ and $G_{ij}(E, \boldsymbol{r})$ being the retarded Green's functions in the absence of a point defect. In the following, we take $V_0/t = 1000$ and compute the Green's functions in momentum space, $G_{ij}(E, \boldsymbol{k}) = [(\epsilon - \mathcal{H}_0(\boldsymbol{k}))^{-1}]_{ij}$, where $\epsilon = E + i\eta$ with a broadening factor $\eta = 10^{-4}$ introduced in the following numerical calculations. Accordingly, the retarded Green's functions in real space are related to $G_{ij}(E, \boldsymbol{k})$ by Fourier transformation,

$$G_{ij}(E, \boldsymbol{r}) = \frac{1}{N} \sum_{\boldsymbol{k}} e^{i\boldsymbol{k} \cdot \boldsymbol{r}} G_{ij}(E, \boldsymbol{k}), \tag{14}$$

where we sum over all discrete $k_x$ and $k_y$ points, $k_x = 4\pi n_x/N_x$, $k_y = 2\pi n_y/\sqrt{3}N_y$ and $n_{x/y} = 0, 1, 2, 3, ..., N_{x/y} - 1$. We investigate how the LDOS changes as the size $N = N_x \times N_y$ evolves from $10^2$ to $10^7$.

## B. Retarded Green's functions around the Dirac points

Here we will elaborate the calculation of the Green's functions in the thermodynamic limit. Near the Dirac points, the Hamiltonian, Eq. (5), be diagonalized by a unitary transformation $\hat{E} = U^{-1}\mathcal{H}_0(\boldsymbol{q})U$, where the unitary matrix $U$ connects the eigenbasis $\Phi^\dagger(\boldsymbol{q}) = \left[\phi_{g-}^\dagger(\boldsymbol{q}), \phi_{q-}^\dagger(\boldsymbol{q}), \phi_{q+}^\dagger(\boldsymbol{q}), \phi_{g+}^\dagger(\boldsymbol{q})\right]$, to the site-basis, $\Psi^\dagger(\boldsymbol{k}) = \left[c_{A1}^\dagger(\boldsymbol{k}), c_{A2}^\dagger(\boldsymbol{k}), c_{B1}^\dagger(\boldsymbol{k}), c_{B2}^\dagger(\boldsymbol{k})\right]$, by

$$\Phi(\boldsymbol{q}) = U^{-1}\Psi(\boldsymbol{q}) \tag{15}$$

$$= \frac{1}{M(\boldsymbol{q})} \begin{pmatrix} -h(\boldsymbol{q}) & -\gamma_1 & \gamma_1 & h^*(\boldsymbol{q}) \\ \gamma_1 \frac{h(\boldsymbol{q})}{|h(\boldsymbol{q})|} & -|h(\boldsymbol{q})| & -|h(\boldsymbol{q})| & \gamma_1 \frac{h^*(\boldsymbol{q})}{|h(\boldsymbol{q})|} \\ -\gamma_1 \frac{h(\boldsymbol{q})}{|h(\boldsymbol{q})|} & |h(\boldsymbol{q})| & -|h(\boldsymbol{q})| & \gamma_1 \frac{h^*(\boldsymbol{q})}{|h(\boldsymbol{q})|} \\ h(\boldsymbol{q}) & \gamma_1 & \gamma_1 & h^*(\boldsymbol{q}) \end{pmatrix} \Psi(\boldsymbol{q}), \tag{16}$$

with $M(\boldsymbol{q}) = \sqrt{2}\sqrt{\gamma_1^2 + |h(\boldsymbol{q})|^2} \approx \sqrt{2}\sqrt{\gamma_1^2 + q^2}$. After this unitary transformation, the electronic Hamiltonian becomes a diagonal energy eigenvalue matrix

$$\hat{E}(\boldsymbol{q}) = \begin{pmatrix} -\gamma_1 & 0 & 0 & 0 \\ 0 & -q^2/\gamma_1 & 0 & 0 \\ 0 & 0 & q^2/\gamma_1 & 0 \\ 0 & 0 & 0 & \gamma_1 \end{pmatrix}. \tag{17}$$

The retarded Green's functions in the eigenbasis can be computed straightforwardly, i.e. $G_{g\pm}(E, \boldsymbol{q}) = -i\int_0^\infty dt e^{i(E+i\eta)t} \langle \phi_{g\pm}(t, \boldsymbol{q})\phi_{g\pm}^\dagger(0, \boldsymbol{q})\rangle_0 = 1/[E \mp \gamma_1 + i\eta]$ and $G_{q\pm}(E, \boldsymbol{q}) = 1/[E \mp (q^2/\gamma_1) + i\eta]$ with a broadening factor $\eta$.

Using the transform matrix between the eigenbasis and the site basis, we represent the retarded Green's functions in the site-basis as

$$G_{A_1A_1}(E,\boldsymbol{q}) = \frac{\gamma_1^2}{M(q)^2}\left[G_{q+}(E,\boldsymbol{q})+G_{q-}(E,\boldsymbol{q})\right]+\frac{|h(\boldsymbol{q})|^2}{M(q)^2}\left[G_{g+}(E,\boldsymbol{q})+G_{g-}(E,\boldsymbol{q})\right], \quad (18)$$

$$G_{B_1B_1}(E,\boldsymbol{q}) = \frac{|h(\boldsymbol{q})|^2}{M(q)^2}\left[G_{q+}(E,\boldsymbol{q})+G_{q-}(E,\boldsymbol{q})\right]+\frac{\gamma_1^2}{M(q)^2}\left[G_{g+}(E,\boldsymbol{q})+G_{g-}(E,\boldsymbol{q})\right], \quad (19)$$

$$G_{A_1B_1}(E,\boldsymbol{q}) = \frac{\gamma_1 h^*(\boldsymbol{q})}{M(q)^2}\left[G_{g+}(E,\boldsymbol{q})-G_{g-}(E,\boldsymbol{q})\right]+\frac{\gamma_1 h^*(\boldsymbol{q})}{M(q)^2}\left[G_{q+}(E,\boldsymbol{q})-G_{q-}(E,\boldsymbol{q})\right], \quad (20)$$

$$G_{A_1A_2}(E,\boldsymbol{q}) = \frac{\gamma_1 h^*(\boldsymbol{q})}{M(q)^2}\left[G_{g+}(E,\boldsymbol{q})+G_{g-}(E,\boldsymbol{q})\right]-\frac{\gamma_1 h^*(\boldsymbol{q})}{M(q)^2}\left[G_{q+}(E,\boldsymbol{q})+G_{q-}(E,\boldsymbol{q})\right], \quad (21)$$

$$G_{B_1A_2}(E,\boldsymbol{q}) = \frac{\gamma_1^2}{M(q)^2}\left[G_{g+}(E,\boldsymbol{q})-G_{g-}(E,\boldsymbol{q})\right]-\frac{|h(\boldsymbol{q})|^2}{M(q)^2}\left[G_{q+}(E,\boldsymbol{q})-G_{q-}(E,\boldsymbol{q})\right], \quad (22)$$

$$G_{A_1B_2}(E,\boldsymbol{q}) = \frac{[h^*(\boldsymbol{q})]^2}{M(q)^2}\left[G_{g+}(E,\boldsymbol{q})-G_{g-}(E,\boldsymbol{q})\right]-\frac{\gamma_1^2\,[h^*(\boldsymbol{q})]^2}{|h(\boldsymbol{q})|^2 M(q)^2}\left[G_{q+}(E,\boldsymbol{q})-G_{q-}(E,\boldsymbol{q})\right]. \quad (23)$$

By employing time-reversal and structure symmetries of a bilayer, we obtain $G_{A_1A_1}=G_{B_2B_2}$, $G_{B_1B_1}=G_{A_2A_2}$, $G_{A_1A_2}=G_{B_1B_2}$ and $G_{A_1B_1}=G_{B_2A_2}$. Integrating all states near the Dirac points within a momentum cutoff $\Lambda$, we can obtain the Green's functions in real space, $G_{ij}(E)=\int_{|\boldsymbol{q}|<\Lambda}\frac{d^2\boldsymbol{q}}{4\pi^2}G_{ij}(E,\boldsymbol{q})$, where $i,j=A_1,B_1,A_2,B_2..$

Near the Dirac points $\boldsymbol{k}=(4\pi/3,0)$, $h(\boldsymbol{k})$ is expanded as

$$h(\boldsymbol{q}) \simeq q\left(\frac{\sqrt{3}}{2}\cos\theta+\frac{1}{8}q\cos^2\theta+\frac{3}{8}q\sin^2\theta\right)+iq\left(-\frac{\sqrt{3}}{2}\sin\theta+\frac{3}{4}q\sin\theta\cos\theta\right), \quad (24)$$

where $\theta$ is the angle centered at $\boldsymbol{k}=(4\pi/3,0)$. We further represent the integral $\int_{|\boldsymbol{q}|<\Lambda}d^2\boldsymbol{q}=\int_0^\Lambda q\,dq\int_0^\pi d\theta$ and compute the angle parts of all Green's functions. To the leading order, we show $\int_0^{2\pi}d\theta h(\boldsymbol{q})=\int_0^{2\pi}d\theta h^*(\boldsymbol{q})\simeq q^2$ and $\int_0^{2\pi}d\theta\,[h(\boldsymbol{q})]^2=\int_0^{2\pi}d\theta\,[h^*(\boldsymbol{q})]^2\simeq 0$. This implies $G_{A_1B_2}(E)\simeq 0$. One can expand $h(\boldsymbol{q})$ to the next order and show $\int_0^{2\pi}d\theta\,[h(\boldsymbol{q})]^2\propto q^6$, which is ignored in our calculation. Using the integral table and taking $\eta\to 0$, we obtain the leading order of the Green's functions as

$$G_{A_1A_1}(E) \simeq \frac{-E}{\gamma_1^2-E^2}\left[\ln(E)^2+2\Lambda^2\right]+i\frac{\gamma_1^2}{|E|-\gamma_1}, \quad (25)$$

$$G_{B_1B_1}(E) \simeq \frac{1}{\gamma_1^2-E^2}E\ln(E)^2+i\frac{\gamma_1|E|}{|E|-\gamma_1}, \quad (26)$$

$$G_{A_1B_1}(E) \simeq -\frac{\Lambda^2}{\gamma_1^2-E^2}+i\,\text{sign}(E)\frac{|E|}{|E|-\gamma_1}, \quad (27)$$

$$G_{A_1A_2}(E) \simeq \frac{-E}{\gamma_1^2-E^2}\left[\ln(E)^2+2\Lambda^2\right]+i\frac{|E|}{\gamma_1-|E|}, \quad (28)$$

$$G_{B_1A_2}(E) \simeq \frac{1}{\gamma_1^2-E^2}E^2\ln(E)^2+i\,\text{sign}(E)\frac{|E|}{\gamma_1-|E|}. \quad (29)$$

We note that LDOS in the pristine BLG can be computed by $\rho_i^0(E) = -\text{Im}\,[G_{ii}(E)]\,/\pi$ where $i = A_1, A_2, B_1, B_2$. It is easy to show $\rho_{A_1}^0(E) = \rho_{B_2}^0(E) \simeq \gamma_1$ and $\rho_{A_2}^0(E) = \rho_{B_1}^0(E) \propto |E|$.

### C.  Integral table

Here we list some useful integrations to compute the Green's functions :

$$\int_0^\Lambda dq \frac{q}{\gamma_1^2 + q^2} = -\frac{1}{2}\ln\left(\frac{\gamma_1^2}{\Lambda^2 + \gamma_1^2}\right), \tag{30}$$

$$\int_0^\Lambda dq \frac{q^3}{\gamma_1^2 + q^2} = \frac{1}{2}\Lambda^2 + \frac{1}{2}\gamma_1^2\ln\left(\frac{\gamma_1^2}{\Lambda^2 + \gamma_1^2}\right), \tag{31}$$

$$\lim_{\eta \to 0}\int_0^\Lambda dq \frac{q^3}{\gamma_1^2 + q^2}\frac{E \mp q^2/\gamma_1}{(E \mp q^2/\gamma_1)^2 + \eta^2} = \frac{\gamma_1}{4(E \pm \gamma_1)}\left\{2\gamma_1\ln\left(\frac{\gamma_1^2}{\Lambda^2 + \gamma_1^2}\right) \pm E\ln\left(\frac{E\gamma_1}{\Lambda^2 \mp E\gamma_1}\right)^2\right\} \tag{32}$$

$$\lim_{\eta \to 0}\int_0^\Lambda dq \frac{q}{\gamma_1^2 + q^2}\frac{E \mp q^2/\gamma_1}{(E \mp q^2/\gamma_1)^2 + \eta^2} = \frac{1}{4(E \pm \gamma_1)}\left\{\ln\left(\frac{E}{\gamma_1}\right)^2 + \ln\left(\frac{\lambda^2 + \gamma_1^2}{\Lambda^2 \mp E\gamma_1}\right)^2\right\}. \tag{33}$$

### D.  Harper equations

The Harper equations of a pristine BLG at $A_1$, $A_2$, $B_1$ and $B_2$ sites can be expressed as

$$\sum_{\boldsymbol{\delta}_i}\varphi_{B_1}(\boldsymbol{r} + \boldsymbol{\delta}_i) + \gamma_3\sum_{\boldsymbol{\delta}_i'}\varphi_{B_2}(\boldsymbol{r} + \boldsymbol{\delta}_i') = E\varphi_{A_1}(\boldsymbol{r}), \tag{34}$$

$$\sum_{\boldsymbol{\delta}_i}\varphi_{B_2}(\boldsymbol{r} + \boldsymbol{\delta}_i) + \gamma_1\varphi_{B_1}(\boldsymbol{r}) = E\varphi_{A_2}(\boldsymbol{r}), \tag{35}$$

$$\sum_{\boldsymbol{\delta}_i'}\varphi_{A_1}(\boldsymbol{r} + \boldsymbol{\delta}_i') + \gamma_1\varphi_{A_2}(\boldsymbol{r}) = E\varphi_{B_1}(\boldsymbol{r}), \tag{36}$$

$$\sum_{\boldsymbol{\delta}_i'}\varphi_{A_2}(\boldsymbol{r} + \boldsymbol{\delta}_i') + \gamma_3\sum_{\boldsymbol{\delta}_i}\varphi_{A_1}(\boldsymbol{r} + \boldsymbol{\delta}_i) = E\varphi_{B_2}(\boldsymbol{r}), \tag{37}$$

where $\boldsymbol{\delta}_i$ and $\boldsymbol{\delta}_i'$ are the three displacement vectors pointed from a $B_2$ site to the nearest neighbors $A_1$ sites on the top layer and $A_2$ sites on the bottom layer respectively.

## V.  ACKNOWLEDGE

We thank Hsiu-Hau Lin for useful discussions. JSY is supported by the Ministry of Science and Technology, Taiwan through grant MOST 104-2917-I-564-054. WMH acknowledge supports from the National Science Council in Taiwan through grant MOST 104-2112-M-005-006-MY3. Financial supports and friendly environment provided by the National Center for Theoretical Sciences in Taiwan are also greatly appreciated.

## VI.  AUTHOR CONTRIBUTIONS

J.S.Y and W.M.H. perform the analytical calculations and the finite-size simulations. J.M.T performs the calculations in the thermodynamics limit. W.M.H. supervises the whole work. All authors contribute to the preparation of this manuscript.

## VII.  ADDITIONAL INFORMATION

**Competing financial interests:** The authors declare no competing interests.